\documentclass[prb,twocolumn,showpacs,preprintnumbers,superscriptaddress]{revtex4}
\usepackage{graphics}
\usepackage{graphicx}
\usepackage{dcolumn} 
\usepackage{amsmath}
\usepackage{bm}
\usepackage{color}
\usepackage{epsfig}
\newcommand{\bnoo}{Ba$_2$NaOsO$_6$}
\newcommand{\byoo}{Ba$_2$YOsO$_6$}

\begin{document}
\title{Calculated g-factors of $5d$ double perovskites \bnoo~ and \byoo
}
\author{Kyo-Hoon Ahn}
\affiliation{Department of Applied Physics, Graduate School, Korea University, Sejong 33019, Korea}

\author{Karel Pajskr}
\affiliation{Charles University in Prague, Faculty of Mathematics and Physics,
Department of Condensed Matter Physics, Ke Karlovu 5, 121 16 Prague 2, Czech Republic}

\author{Kwan-Woo Lee}
\email{mckwan@korea.ac.kr}
\affiliation{Department of Applied Physics, Graduate School, Korea University, Sejong 33019, Korea}
\affiliation{Department of Display and Semiconductor Physics, Korea University, Sejong 33019, Korea}

\author{Jan Kune\v{s}}
\email{kunes@fzu.cz}
\affiliation{Institute of  Solid State Physics, TU Wien, Wiedner Hauptstr. 8, 1040 Wien, Austria}
\affiliation{Institute of Physics, Academy of Sciences of the Czech Republic, Cukrovarnick\'a 10,
Praha 6, 162 53, Czech Republic}

\begin{abstract} 
Using Wannier functions to represent the density functional results we calculate the hybridization corrections
to the orbital momentum operator in the Os $5d$ shell of Mott insulators \bnoo~ and \byoo.
The $g$-factors are obtained by evaluating the spin and orbital momentum operators in the
atomic ground states of the Os ion. While the hybridization corrections play minor role in 
$d^3$ ion of \byoo~ with dominant spin moment, they are instrumental for observation
of non-zero $g$-factor of the $d^1$ ions of \bnoo. In addition we analyze the exchange interactions
in \byoo~ and find them consistent with the reported magnetic structure.
\end{abstract}

\date{\today}

\maketitle

\section{Introduction}
The high temperature paramagnetic moment is one of the basic characteristics of magnetic materials
which is often used to draw conclusions about  the underlying atomic multiplets. 
In Mott insulators the charge fluctuations are quenched and the low-energy physics is usually
governed by a single atomic multiplet. Its symmetry properties (degeneracy), captured by pseudospin description,
are determined by the gross features of the ligand field and do not depend on the material details. 
The size of the local magnetic moment, encoded in the $g$-factor, is affected by hybridization with ligands
which modifies the shape of the local orbitals. This is particularly important
for heavier $d^1$ ions in strong cubic crystal-field. The spin-orbit interaction (SOI) selects the $J_{\text{eff}}=3/2$ quadruplet to 
be the ground state. For purely $d$ orbitals the spin and orbital contributions to the magnetic moment within the 
$J_{\text{eff}}=3/2$ subspace exactly cancel each other leading to $g=0$.~\cite{abragam}
Magnetic response of such ion can therefore arise only due to mixing with higher lying multiplets or hybridization
correction to the $g$-factor.
\bnoo~is a rare example of a $d^1$ insulators with cubic symmetry.

In this article, we present numerical calculation of $g$-factors based on the Wannier orbitals 
and diagonalization of the atomic problem. It is applied to two
$5d$ double-perovskite insulators \bnoo~(BNOO) and \byoo~(BYOO) that attracted some interest recently.\cite{LP07,pickett16}

\section{Numerical Approach}
\subsection{Electronic structure calculations}
First we carried out {\it ab initio} calculations within the the generalized gradient approximation~\cite{gga}
to the density functional theory using the full-potential code {\sc wien2k}.~\cite{wien1,wien2}
We studied three compounds:  the cubic perovskite SrVO$_3$ (SVO) as a simple model system to
debug our codes and two cubic double perovskites BNOO and BYOO as the actual goals of the present work.
The following lattice parameters were used: $a=$3.8425 \AA~ for SVO\cite{svo}
and $a=$8.2870 (8.5341) with the oxygen internal parameter $x=$0.2272 (0.2350) for BNOO (BYOO)\cite{bnoo,byoo}.
The SOI was included for the $5d$ systems of BNOO and BYOO,
while not for the $3d$ system SVO.

The {\sc wien2k} bases were characterized by $R_{mt}K_{max} = 7$ 
and muffin-tin spheres with radii in atomic units: Sr (2.5), V (1.91), and O (1.70) for SVO;
Ba (2.5), Y (2.18), Os (2.03), and O (1.66) for BYOO;
Ba (2.5), Na (2.15), Os (1.96), and O (1.61) for BNOO.

\begin{figure}[tbp]
\includegraphics[scale=0.18]{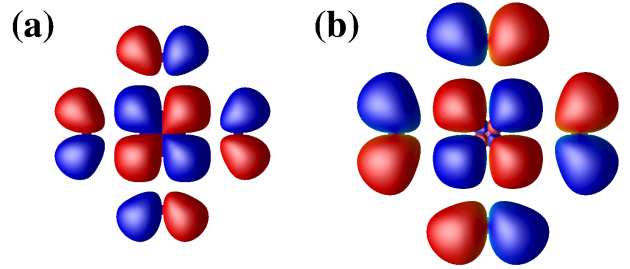}
\caption{(color online)  $d_{xy}$ WF plot of (a) SVO and (b) BNOO.
The latter is identical to that of BYOO.
The isovalues are 2.0 for SVO and 0.5 for BNOO.
}
\label{fig1}
\end{figure}

\begin{figure}[tbp]
\includegraphics[scale=0.35]{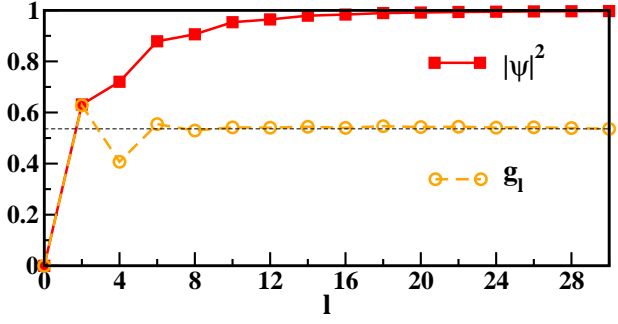}
\caption{(color online)  Convergence of the weight of the Wannier function ($|\psi|^2$)  
and the one-particle orbital $g$-factor ($g_l$)
with the spherical harmonic order $l$ for BNOO.
Note that there is no contribution of the odd $l$ terms because of the WF parity.
These data are nearly identical for each $t_{2g}$ orbital of BNOO and BYOO. 
}
\label{fig2}
\end{figure}

\subsection{Evaluation of angular and spin momenta}
While spin magnetization is straightforward to calculate, orbital magnetization is a more complicated problem. Ceresoli {\it et al.}~\cite{ceresoli06}
provided a general basis independent expression for orbital magnetization of a non-interacting insulator in terms of Wannier functions (WFs),
which resolved the problem that orbital currents cannot, in general, be assigned to a particular atomic site. The Mott insulating nature of the studied 
materials simplifies our task considerably. We assume that the orbital current density is localized in regions around the active (Os) atoms separated
by regions where it vanishes. Working in the basis of atom-centered Wannier functions we can express the orbital magnetization as a sum of
individual atomic contributions.

The WFs~\cite{wannier37,marzari12} with the $|d_{xy},\uparrow\rangle$, $|d_{yz},\uparrow\rangle$, $|d_{xz},\uparrow\rangle$, $|d_{xy},\downarrow\rangle$, $|d_{yz},\downarrow\rangle$, and $|d_{xz},\downarrow\rangle$ symmetry spanning the isolated $t_{2g}$ manifold 
were obtained using the {\sc wannier90}\cite{wannier90} and {\sc wien2wannier}\cite{wannier} codes.
Without SOI these WFs are well defined by their symmetry properties. With SOI this in general is not the case due to mixing of the spin species.
However, since the effect of SOI is predominantly localized on the $5d$ ion, 
it is possible to construct the WFs which are to high accuracy eigenstates
of the spin $S_z$ operator. The on-site one-particle Hamiltonian in this $t_{2g}$ Wannier basis has the textbook form.

We have used $8\times8\times8$ $k$-mesh for SVO and $4\times 4\times 4$ $k$-mesh for BNOO and BYOO, which
ensured that the WFs were spatially well separated from their aliases.
The normalized WFs $\psi_{\alpha}(\mathbf{r})$ were expanded in spherical harmonics
inside a sphere of with radius $4a$ ($\sqrt{2}a$) around the V (Os) atom for SVO (BNOO,BYOO) . 
\begin{eqnarray}
\psi_{\alpha}(\mathbf{r})=\sum_{l,m}c_{lm}^{\alpha}(r)Y_{lm}(\hat{\mathbf{r}}).
\label{eq1}
\end{eqnarray}

To this end we have employed a dense spherical mesh of 100$\times$60$\times$120
and 400$\times$180$\times$360 of $(r,\theta,\phi)$ points for SVO and BNOO/BYOO, respectively.
The maximum values of $l$ for this expansion were $l_{max}$=16 for SVO and 
30 for BNOO and BYOO, which were sufficient to capture 99\% of the WF.


The spherical harmonic representation allows straightforward evaluation of the 
angular momentum matrix elements in the $t_{2g}$ subspace $\mathbf{l}_{\alpha\beta}$
\begin{equation}
\begin{split}
&l^z_{\alpha\beta}\equiv<\psi_{\alpha}|\hat{l}^{z}|\psi_{\beta}>=\sum_{l,m} m(c_{lm}^{\alpha}|c_{lm}^{\beta})\\
&l^+_{\alpha\beta}=\sum_{l,m}\sqrt{l(l+1)-m(m+1)}(c_{lm+1}^{\alpha}|c_{lm}^{\beta})\\
&l^x_{\alpha\beta}=
\frac{l^+_{\alpha\beta}+(l^+_{\alpha\beta})^{\dagger}}{2},
l^y_{\alpha\beta}=
\frac{l^+_{\alpha\beta}-(l^+_{\alpha\beta})^{\dagger}}{2i}.
\label{eq2}
\end{split}
\end{equation}
Here, $(f_{1}|f_{2})$ stands for
the radial integral $(f_{1}|f_{2})=\int_{0}^{\infty}dr\ r^{2}\ f_{1}^{*}(r)f_{2}(r)$.
For the cubic site symmetry, $\mathbf{l}$ has the same form as for pure $d$ orbitals ($\mathbf{l}_d$). 
The higher harmonics in the site expansion of WFs introduce a multiplicative correction factor
$\mathbf{l}=\gamma\mathbf{l}_d$. The physical origin of $\gamma$ is the hybridization with ligands.
Using matrix elements of Eq. (\ref{eq2}) we can write the orbital momentum operator 
in the $t_{2g}$ subspace
\begin{equation}
\hat{\mathbf{l}}=\sum_{m,n}\mathbf{l}_{mn}c^{\dagger}_{m}c^{\phantom\dagger}_{n}
\end{equation}
with $m\equiv k\alpha$ being spin/orbital indices. There is no hybridization
screening of the spin moment and thus the spin momentum operator has the standard form
\begin{equation}
\hat{\mathbf{s}}=\frac{1}{2}\sum_{k,\alpha,\beta}\boldsymbol{\sigma}_{\alpha\beta}c^{\dagger}_{k\alpha}c^{\phantom\dagger}_{k\beta}.
\end{equation}

As the final step, we diagonalize the on-site Hamiltonian including the on-site Coulomb interaction and SOI. 
We denote the matrix elements of the orbital $\mathbf{\hat{l}}$ and spin $\mathbf{\hat{s}}$ momenta within the ground state multiplet
with capital $\mathbf{L}$ and $\mathbf{S}$, respectively.
Comparing of the magnetic moment elements  $\mathbf{M}=\mu_B(2\mathbf{S}+\mathbf{L})$ 
to the ground state pseudospin operator $\mathbf{J}$, 
we obtain the $g$-factor and the effective paramagnetic moment $\mu_{\text{eff}}$.

\section{Results and discussion}


\subsection{Test case: SrVO$_3$}
To test our implementation of Eq. (\ref{eq2}) we have chosen a simple system without SOI. Since SrVO$_3$ is 
a metal, we do not assign a physical significance to this result. We merely want to show that $\mathbf{l}$
has the expected form. Evaluating Eq. (\ref{eq2}) yields
\begin{eqnarray}
\bm{l_x}= i
\left(\begin{array}{ccc}
 0.0000 &  0.0000 & -0.7955 \\
 0.0000 &  0.0000 &  0.0000 \\
 0.7955 &  0.0000 &  0.0000 \\
\end{array}\right)\\
\bm{l_y}= i
\left(\begin{array}{ccc}
 0.0000 &  0.7955 &  0.0000 \\
-0.7955 &  0.0000 &  0.0000 \\
 0.0000 &  0.0000 &  0.0000 \\
\end{array}\right)\\
\bm{l_z}= i
\left(\begin{array}{ccc}
 0.0000 &  0.0000 &  0.0000 \\
 0.0000 &  0.0000 &  0.7956 \\
 0.0000 & -0.7956 &  0.0000 \\
\end{array}\right),
\label{eq3}
\end{eqnarray}
which amounts to $\gamma=0.7956$.


\subsection{$d^3$ case: \byoo}
The structural and magnetic properties of BYOO were recently investigated by Kermarrec~{\it et al.}~\cite{byoo}, who reported a frustrated
A-type antiferromagnet (AFM) with the N\'eel temperature $T_N\sim$~69~K, Curie-Weiss temperature $\Theta=717$~K, 
and the effective paramagnetic moment
$\mu_{\text{eff}}=3.91$~$\mu_B$ close to the spin only value of 3.87~$\mu_B$.

Using the above procedure we have obtained $\gamma$ of 0.587 for Wannier functions 
built from the subspace spanning the Os $5d$ bands
($d$-only model). Next, we have used the on-site part of the tight-binding Hamiltonian (including SOI) augmented with the atomic-like Coulomb interaction, parameterized
with $U=1.8$~eV and $J=0.4$~eV~\cite{pajskr16}, and diagonalized it. The fourfold degenerate ground state can be represented as $J=3/2$ pseudospin. 
Calculating the matrix elements of the magnetic moment $\mathbf{M}=\mu_B(2\mathbf{S}+\mathbf{L})$ 
in the ground state multiplet, 
we arrive at the $g$-factor of 1.89, leading to $\mu_{\text{eff}}=3.66\mu_B$.
The matrix elements of $\mathbf{S_z}$ and $\mathbf{L}_z$ are shown in the Appendix.
This shows that even the relatively strong SOI leads to only a small correction
in case of a half-filled $5d$ ($t_{2g}$) shell and that the local moment is dominated by the spin component.\cite{naoso3}
Interestingly, under special circumstances the small corrections may play an important role. Such is the case of
the ferrimagnet Sr$_2$CrOsO$_6$. Without SOI the band structure calculations~\cite{lee08,das11} show complete 
compensation of spin moments of Cr and Os and only the uncompensated orbital moments arising due to 
SOI give rise to a net magnetization.

Next, we have calculated the effective interaction between the $J=3/2$ pseudospins 
on the nearest- and next-nearest-neighbor (NNN) bonds. The spin-orbit entangled atomic states, in general, 
lead to an anisotropic exchange with possibly strong terms beyond dipole-dipole interaction and may
give rise to exotic magnetic orders.~\cite{chen10,pi14}
A full $16\times16$ interaction matrix was calculated using the Schrieffer-Wolff transformation~\cite{schrieffer66} (`$t^2/U$'-expansion) 
and expanded in spherical tensors. Consistent with the dominant spin nature of the pseudospins
we find that the isotropic dipole-dipole exchange dominates by more than order of magnitude. 
A small anisotropy was found between $J_{\parallel}=2.96$~meV and $J_{\perp}=3.13$~meV 
the interaction of components in and out-of the cubic plane for a given NN bond. The NNN interactions are about 30 times weaker than the NN ones.
The dominant NN AFM interaction is consistent with the A-type ordering. 
We also point out the exchange anisotropy favors the ordered 
moments to lie in the ferromagnetic (FM) planes. 
However, the observed anisotropy is rather small to draw solid conclusions without detailed calculations, which 
are beyond the scope of the present study. The estimated Curie-Weiss temperature $\Theta=12J_{\text{NN}}\simeq 420$~K is somewhat smaller
than the experimental value. Both the theoretical and experimental values contain uncertainties. On the theoretical side it is the value of the on-site Coulomb
interaction $U$ and the limitations of the perturbative expansion. On the experimental side it is in particular the temperature range
of the study which is substantially below the experimental $\Theta$ as pointed by authors of Ref.~[\onlinecite{byoo}].

\subsection{$d^1$ case: \bnoo}
BNOO is an unusual example of $d^1$ Mott insulator.  
It exhibits a FM polarization below 10~K.\cite{LP07,erickson07} 
The crystal and magnetic structures are still under debate.
Specific heat measurements indicate that only $\ln 2$ entropy is released above the transition temperature which contradicts the fourfold degenerate
$J=3/2$ ground state in cubic site symmetry. The structural measurements, nevertheless, indicate that a potential non-cubic distortion must be rather small.
In the following, we address the question of the origin of the high temperature paramagnetic response and the size of the $g$-factor. This is motivated 
by an observation that the magnetic moment $\mathbf{M}=\mu_B(2\mathbf{S}+\mathbf{L})$ has only zero matrix elements within the 
ground state $J=3/2$ multiplet of $d^1$ system.~\cite{abragam} In other words, in the $J=3/2$ quadruplet the spin and orbital contributions to the magnetic moment
exactly compensate each other. 
This conclusion remains valid also in case of weak non-cubic perturbations as long as the mixing with 
excited $J=1/2$ states can be neglected. The likely explanation of the paramagnetic moment is the fact that the local orbitals (Wannier functions) do not have
the exact $d$-symmetry, but contain also higher harmonics which rescale the orbital momentum by $\gamma$.

The $d$--$d$ electron-electron interaction is not effective in the $d^1$ case and thus we only need to consider the SOI. 
The relationship between
$g$ and $\gamma$  in the $J=3/2$ ground state is given by $g=\tfrac{2}{3}(1-\gamma)$.
Our calculations yield 
$\gamma=0.536$ leading to $g=0.31$ and $\mu_{\text{eff}}\simeq0.60\mu_B$. The calculated matrix elements
$\mathbf{l}$ as well as $\mathbf{S}_z$ and $\mathbf{L}_z$ are shown in the Appendix.
This value compares well to the experimentally reported values of 0.596-0.647~\cite{erickson07} and 0.677~\cite{bnoo}.

Another contribution to the paramagnetic response is the van Vleck susceptibility arising from field-induced mixing of the excited states. Since this contribution
is temperature independent, it is experimentally easy to distinguish from the Curie-Weiss response of local moments. A straightforward calculation
in the present atomic treatment yields van Vleck susceptibility $\chi_{\text{vV}}\text{[emu/mol]}\simeq 3.232\times 10^{-5}/\Delta\text{[eV]}$ inversely
proportional to the spin-orbit splitting $\Delta\simeq$~0.49~eV of the $t_{2g}$ shell. The resulting number compares well to the temperature independent 
contribution to the susceptibility ($3\times 10^{-5}$ - $1.8\times 10^{-4}$ emu/mol) reported in Ref.~\onlinecite{erickson07}.

\section{Conclusions}
We have presented a first principles calculation of the $g$-factors of $5d$ ions in selected double perovskite Mott insulators. The calculated effective paramagnetic 
moments $\mu_{\text{eff}}$ compare well to their experimental counterpart obtained from Curie-Weiss fits to the high temperature susceptibility.
We find that even in the case of $5d$ electrons with relatively strong SOI the half-filled shell $d^3$ configuration is dominated by the first Hund's rule
and the local moments have predominantly spin character. In $d^1$ configuration the spin and orbital contributions to $\mu_{\text{eff}}$ 
cancel each other for orbitals with pure $d$ character. The admixture of higher harmonics to the $d$-like Wannier functions due to the hybridization with ligands
partially lifts this exact cancelation and gives rise to small $\mu_{\text{eff}}$ observed in experiments.



\section{Acknowledgments}
This research at KU was supported by NRF of Korea Grant No. NRF-2016R1A2B4009579.
J.K. and K.-H. A. acknowledge the support of the Czech Science Foundation through Project No. 13-25251S.

\section{Appendix}
The matrix elements of orbital momentum operator $\hat{\mathbf{l}}$ in $5d$ shell of Os in BNOO in the basis
of $\{|d_{xy},\uparrow>$, $|d_{yz},\uparrow>$, $|d_{xz},\uparrow>$, $|d_{xy},\downarrow>$, $|d_{yz},\downarrow>$, and $|d_{xz},\downarrow>\}$ WFs
calculated according to Eqs. (\ref{eq1}) and (\ref{eq2}). 
\begin{eqnarray*}
\bm{l_x}&=&
\left(\begin{array}{cccccc}
  0.000  &  0.013  & -0.536i &  0.008  &  0.000  &  0.000  \\
  0.013  &  0.000  &  0.000  &  0.000  &  0.041  & -0.013i \\
  0.536i &  0.000  &  0.000  &  0.000  & -0.013i &  0.008  \\
  0.008  &  0.000  &  0.000  &  0.000  & -0.013  & -0.536i \\
  0.000  &  0.041  &  0.013i & -0.013  &  0.000  &  0.000  \\
  0.000  &  0.013i &  0.008  &  0.536i &  0.000  &  0.000  \\
\end{array}\right),\\
\bm{l_y}&=&
\left(\begin{array}{cccccc}
  0.000  & -0.536i & -0.013  & -0.008i &  0.000  &  0.000  \\
  0.536i &  0.000  &  0.000  &  0.000  & -0.008i &  0.013  \\
 -0.013  &  0.000  &  0.000  &  0.000  &  0.013  & -0.041i \\
  0.008i &  0.000  &  0.000  &  0.000  & -0.536i &  0.013  \\
  0.000  &  0.008i &  0.013  &  0.536i &  0.000  &  0.000  \\
  0.000  &  0.013  &  0.041i &  0.013  &  0.000  &  0.000  \\
\end{array}\right),\\
\bm{l_z}&=&
\left(\begin{array}{cccccc}
  0.041  &  0.000  &  0.000  &  0.000  &  0.014  &  0.014i \\
  0.000  &  0.008  & -0.536i &  0.014  &  0.000  &  0.000  \\
  0.000  &  0.536i &  0.008  &  0.014i &  0.000  &  0.000  \\
  0.000  &  0.014  & -0.014i & -0.041  &  0.000  &  0.000  \\
  0.014  &  0.000  &  0.000  &  0.000  & -0.008  & -0.536i \\
 -0.014i &  0.000  &  0.000  &  0.000  &  0.536i & -0.008  \\
\end{array}\right).
\label{eq5}
\end{eqnarray*}
Due to the SOI the spin projection and $t_{2g}$ character are only approximately conserved. This gives
rise to the additional matrix elements. These are, nevertheless, more than order of magnitude smaller
than the dominant terms, and not considered in the evaluation of the $g$-factor.
The results for BYOO show a similar picture.

The matrix elements of the spin and orbital momentum operator in the 
$J=3/2$ ground state of BYOO ($d^3$).
The basis functions are indexed by 
$J_z=3/2,1/2,-1/2,-3/2$.
\begin{eqnarray*}
\bm{S_z}&=&
\left(\begin{array}{cccc}
  1.442  &  0.000  &  0.000 &  0.000   \\
  0.000  &  0.481  &  0.000  &  0.000  \\
  0.000 &  0.000  &  -0.481  &  0.000    \\
  0.000  &  0.000  &  0.000  &  -1.442  \\
\end{array}\right),\\
\bm{L_z}&=&0.587\times
\left(\begin{array}{cccc}
  -0.058  &  0.000  &  0.000 &  0.000   \\
  0.000  &  -0.019  &  0.000  &  0.000  \\
  0.000 &  0.000  &  0.019  &  0.000    \\
  0.000  &  0.000  &  0.000  &  0.058  \\
 \end{array}\right).\\
\label{eq6}
\end{eqnarray*}

The matrix elements of the spin and orbital momentum operator in the 
$J=3/2$ ground state of BNOO ($d^1$).
The basis functions are indexed by $J_z=3/2,1/2,-1/2,-3/2$. Note that when
$e_g$--$t_{2g}$ mixing can be neglected (this case) the basis functions
are completely specified by symmetry.
\begin{eqnarray*}
\bm{S_z}&=&
\left(\begin{array}{cccc}
  \tfrac{1}{2}  &  0  &  0 &  0   \\
  0  &  \tfrac{1}{6}  &  0  &  0  \\
  0 &  0  &  -\tfrac{1}{6}  &  0    \\
  0  &  0  &  0  &  -\tfrac{1}{2} \\
\end{array}\right),\\
\bm{L_z}&=&0.536\times
\left(\begin{array}{cccc}
  -1  &  0  &  0 &  0   \\
  0  &  -\tfrac{1}{3}  &  0  &  0  \\
  0 &  0  &  \tfrac{1}{3}  &  0    \\
  0  &  0  &  0  &  1  \\
 \end{array}\right).\\
\label{eq6}
\end{eqnarray*}


\begin{thebibliography}{10}

\bibitem{abragam} A. Abragam and B. Bleaney, Electron Paramagnetic Resonance of Transition Ions, (Oxford University Press, Oxford, 1970) p. 421.

\bibitem{LP07} K.-W. Lee and W. E. Pickett, 
   Europhys. Lett. {\bf 80}, 37008 (2007).

\bibitem{pickett16} S. Gangopadhyay and W. E. Pickett,  
   Phys. Rev. B {\bf 93}, 155126 (2016); Phys. Rev. B {\bf 91}, 045133 (2015).
   
 \bibitem{ishizuka15} H. Ishizuka and L. Balents, Phys. Rev. B {\bf 92}, 020411 (2015); Phys. Rev. B {\bf 90}, 184422 (2014).

\bibitem{gga} J. P. Perdew, K. Burke, and M. Ernzerhof,   
  Phys. Rev. Lett. {\bf 77}, 3865 (1996).

\bibitem{wien1} K. Schwarz, and P. Blaha, 
   Comput. Mater. Sci. {\bf 28}, 259 (2003).

\bibitem{wien2} P. Blaha, K. Schwarz, G. K. H. Madsen, D. Kvasnicka, and J Luitz, 
  WIEN2k: an Augmented Plane Wave + Local Orbitals Program for Calculating Crystal Properties 
 (Austria: Karlheinz Schwarz, Techn. Universitaet Wien) (2001).


\bibitem{svo} Y. C. Lan, X. L. Chen, and M. He, 
  J. Alloys Comp. {\bf 354}, 95 (2003).

\bibitem{byoo} E. Kermarrec, C. A. Marjerrison, C. M. Thompson, D. D. Maharaj, K. Levin, 
 S. Kroeker, G. E. Granroth, R. Flacau, Z. Yamani, J. E. Greedan, and B. D. Gaulin,
 Phys. Rev. B {\bf 91}, 075133 (2015). 

\bibitem{bnoo} K. E. Stitzer, M. D. Smith, and H.-C. zur Loye, 
 Solid State Sci. {\bf 4}, 311 (2002).

\bibitem{ceresoli06} D. Ceresoli, T. Thonhauser, D. Vanderbilt, and R. Resta, Phys. Rev. B {\bf 74}, 024408 (2006).

\bibitem{wannier37} G. H. Wannier, Phys. Rev. {\bf 52}, 191 (1937).

\bibitem{marzari12} N. Marzari, A. A. Mostofi, J. R. Yates, I. Souza, and D. Vanderbilt, Rev. Mod. Phys. {\bf 84}, 1419 (2012).

\bibitem{wannier90} A. A. Mostofi, J. R. Yates, Y.-S. Lee, I. Souza, D. Vanderbilt, and N. Marzari, 
 Comput. Phys. Commun. {\bf 178}, 685 (2008).

\bibitem{wannier} J. Kune\v{s}, R. Arita, P. Wissgott, A. Toschi, H. Ikeda, and K. Held, 
 Comput. Phys. Commun. {\bf 181}, 1888 (2010).


\bibitem{pajskr16} K. Pajskr, P. Nov\'ak, V. Pokorn\'y, J. Koloren\v{c}, R. Arita, and J. Kune\v{s}
Phys. Rev. B {\bf 93}, 035129 (2016).

\bibitem{lee08} K.-W. Lee and W. E. Pickett, Phys. Rev. B {\bf 77}, 115101 (2008).

\bibitem{das11} H. Das, P. Sanyal, T. Saha-Dasgupta, and D. D. Sarma, Phys. Rev. B {\bf 83}, 104418 (2011).

\bibitem{chen10} G. Chen, R. Pereira and L. Balents, Phys. Rev. B {\bf 82}, 174440 (2010).

\bibitem{pi14} S.-T. Pi, R. Nanguneri and S. Savrasov, Phys. Rev. Lett. {\bf 112}, 077203 (2014).

\bibitem{schrieffer66} J. R. Schrieffer and P. A. Wolff, Phys. Rev. {\bf 149}, 491 (1966).
\bibitem{naoso3} M.-C. Jung, Y.-J. Song, K.-W. Lee, and W. E. Pickett,
  Phys. Rev. B {\bf 87}, 115119 (2013).

\bibitem{erickson07} A. S. Erickson, S. Misra, G. J. Miller, R. R. Gupta, Z. Schlesinger, 
 W. A. Harrison, J. M. Kim, and I. R. Fisher
 Phys. Rev. Lett. {\bf 99}, 016404 (2007).


\end{thebibliography}
\end{document}